# Helicity resolved Raman scattering of $MoS_2$, $MoSe_2$, $WS_2$ and $WSe_2$ atomic layers

*Shao-Yu Chen[†], Changxi Zheng[‡§], Michael S. Fuhrer[‡] and Jun Yan[†\*]*

[†]Department of Physics, University of Massachusetts, Amherst, Massachusetts 01003, USA

[‡]School of Physics and [§]Department of Civil Engineering, Monash University, Victoria 3800, Australia

[\*]Corresponding Author: Jun Yan.   Tel:   (413)545-0853   Fax:   (413)545-1691   E-mail: yan@physics.umass.edu

**Abstract**

The two-fold valley degeneracy in two-dimensional (2D) semiconducting transition metal dichalcogenides (TMDCs) $(Mo,W)(S,Se)_2$ is suitable for "valleytronics", the storage and manipulation of information utilizing the valley degree of freedom. The conservation of luminescent photon helicity in these 2D crystal monolayers has been widely regarded as a benchmark indicator for charge carrier valley polarization. Here we perform helicity-resolved Raman scattering of the TMDC atomic layers. In drastic contrast to luminescence, the dominant first-order zone-center Raman bands, including the low energy breathing and shear modes as well as the higher energy optical phonons, are found to either maintain or completely switch the helicity of incident photons. These experimental observations, in addition to providing a useful tool for characterization of TMDC atomic layers, shed new light on the connection between photon helicity and valley polarization.

**Manuscript text**

Since the discovery of graphene,[1] the mechanical, electronic, chemical and optical properties of various two-dimensional (2D) materials as well as their heterostructures have been widely investigated.[2-5] A prominent example is the semiconducting transition metal dichalcogenides (TMDCs) that exhibit rich physical phenomena, including indirect to direct bandgap transition,[6, 7] large exciton and trion binding energy,[8-11] strong photoluminescence and electroluminescence,[7, 12-14] superior transistor performance with large on-off ratio[15-17] and reasonably high mobility,[5, 18, 19] and perhaps most strikingly, the capability to address the valley degree of freedom.[20-24] Manipulation of valley polarized carriers excited by circularly polarized light has led to recent observation of the valley Hall effect[25] that opens up potential for applications in 'valleytronics' envisioned before in graphene.[26, 27] Here we apply circularly polarized light to excite electrons in TMDC atomic layers and measure the helicity of the photons emitted after they are inelastically scattered by phonons. We discovered that while some phonons maintain helicity from incident to emitted photon, others can switch it completely. Our results can be explained by the symmetry of participating lattice vibrations in the Raman scattering process. The helicity selection rules provide clean Raman spectra and prove to be a powerful tool for resolving phonon mode assignment and characterization. Importantly, the helicity of Raman scattered photons is independent of layer number and excitation laser wavelength, drastically different from observations in valley pumping of TMDC with helicity resolved luminescence. Our experiments consequently provide new insights into the relation between the photon helicity and valleytronics in semiconducting TMDCs.

Layered TMDC materials have a graphite-like structure with each graphene sheet replaced with an X-M-X or $MX_2$ tri-atomic layer, where X is a chalcogen atom (e.g. sulfur, selenium, tellurium) and M is a transition metal atom (e.g. tungsten, molybdenum). One layer $MX_2$ ($1L-MX_2$) has three atoms in its unit cell and their vibrations result in 9 normal modes, including 3 acoustic and 6 optical branches: one each for LA, TA, ZA and two each for LO, TO, ZO (the LA, TA and LO, TO are longitudinal and transverse in-plane acoustic and optical modes, while ZA and ZO are out-of-plane vibrations, following the name convention established in graphene[28] and graphite[29] research for layered 2D materials). Figure 1(a) illustrates the two types of optical phonons at the Brillouin zone center (the Γ point): one type involves only relative motion of the two chalcogen atoms, and the other type is participated by both the transition metal and the chalcogen atoms. To facilitate discussion, we shall call the in-plane chalcogen vibration the 'IC' mode, the in-plane relative motion of transition metal and chalcogen atoms the 'IMC' mode. Similarly the out-of-plane phonon involving only chalcogen atoms is called the 'OC' mode, and the out-of-plane phonon involving both chalcogen and transition metal atoms is named the 'OMC' mode. Similar to graphene, the in-plane IC and IMC optical phonons are each doubly degenerate at the Brillouin zone center, in connection with the LO and TO branches at finite wavevectors.

For multi-layer $MX_2$, the optical phonons within individual $MX_2$ layer couple to each other and create new normal modes. Due to interlayer interactions such as van der Waals force and dielectric screening, energies of the new modes are slightly shifted from the corresponding monolayer phonon. Similarly, the acoustic phonons of individual layers couple to each other and form new optical phonon branches, including the in-plane

LO/TO shear mode and the out-of-plane ZO breathing mode (see Fig. 1(b) for the case of Hc stacking[30] 2L-$MX_2$). These modes have relatively low energy reflecting that the interlayer interactions are much weaker than the covalent/ionic bonding within each $MX_2$ layer.

Our studies focus on these optical phonons in the four prototypical semiconducting TMDCs including $MoS_2$, $WS_2$, $MoSe_2$ and $WSe_2$. These samples (optical microscope images in Fig. 1(c-f)) are either exfoliated from bulk or grown by chemical vapor deposition (CVD, see Methods). Figure 1(g) shows a schematic drawing of our experimental setup. We excite TMDC atomic layers with σ+ circularly polarized laser light and detect separately the σ+ and σ- luminescent and Raman signals using a triple spectrometer equipped with a liquid nitrogen cooled CCD (details in Methods). While there have been many works that measure unpolarized or linearly polarized TMDC Raman spectra,[31-40] and circularly polarized light has been used for Raman studies of, e.g., graphene, $CaC_6$ and cuprates,[41-43] the measurement here is the first that systematically analyses the helicity of Raman signal emitted by TMDC atomic layers.

We first verify our setup by measuring the photoluminescence of monolayer $MoS_2$. Figure 2(a) shows that the helicity of photons depends sensitively on the incident photon energy, consistent with previous measurements on $MoS_2$ and other TMDCs.[21-24] The dominance of luminescence in the σ+ polarization with 1.92 eV σ+ excitation has been ascribed as a consequence of charge carrier valley pseudospin conservation, and the un-polarized emission with 2.54 eV excitation has been interpreted as an indication of loss of valley polarization. These studies have unveiled the rich valley physics of TMDCs arising from the broken inversion symmetry and strong spin-orbit interaction.[20]

Using the same scattering geometry we measured the photon helicity-resolved Raman spectra of $MoS_2$. As shown in Fig. 2(b), the OC phonon scattered photons have the same σ+ helicity as the incident photon while the IMC phonon band is observed only in the opposite σ- polarization. For further confirmation, we rotated the half-waveplate in the collection path and measured the angular dependence of the spectra. The normalized intensities of the IMC (blue), OC (orange) and Rayleigh (green) scattered light of 1L-$MoS_2$ are plotted as a function of detection angle in the inset of Fig. 2(b). The OC scattered photon has exactly the same polarization as Rayleigh scattered light, while IMC scattered photon has perfectly opposite polarization. To test the robustness of the observed helicity selectivity, we examined the helicity of the scattered photons in multilayers and with different laser excitations (Argon, Nd:YAG, and dye laser; see Methods). As shown in Fig. 2(b) and Fig. 2(c), the photon helicity is independent of layer number and incident photon energy: the IMC mode always switches the photon helicity. This is in drastic contrast to luminescence spectra in Fig. 2(a) where the percentage of photons that maintain the incident photon polarization depends sensitively on incident photon energy as well as TMDC layer numbers.[21] Furthermore for luminescence σ- intensity never exceeds that of σ+ at any photon energy.

At least for some incident photon energies (e.g. 1.95 eV), a high degree of valley polarization of the excited carriers is anticipated. An interesting question then arises: is the IMC mode switching the valley index of photo-excited carrier, i.e., is there inter-valley scattering during this Raman process? To understand this we first note that the IMC is a Brillouin zone center phonon, while inter-valley scattering requires large momentum transfer (~$10^8$ cm$^{-1}$) that is three orders of magnitude larger than the photon momentum (~$10^5$ cm$^{-1}$). Thus, conservation of momentum dictates that the IMC Raman process can only occur within the same valley, despite that the photon helicity is switched. Our data consequently point to the conclusion that in semiconducting TMDCs, even for the monolayer, photons emitted from the *same* valley can have either σ+ or σ- polarization, and the valley-photon helicity selection rule can only be approximately true.

The connection between the valley degree of freedom and the photon helicity in 1L-MX$_2$ is based on angular momentum conservation. For electron states near the corner K+ and K- points there are three contributions to angular momentum, from the spin, the atomic orbit and the lattice.[20] Optical excitation of excitons in the two valleys requires angular momentum transfer of $\pm\hbar$ only when at the K points and when the conduction (valence) band is purely composed of $d_{z^2}$ ($\frac{1}{\sqrt{2}}[d_{x^2-y^2} \pm id_{xy}]$) orbitals. This is however, not exactly true. Tight binding and density functional theory calculations show that the electron wavefunctions do have finite albeit small contributions from the *s* and *p* orbitals of the chalcogen atoms.[44-46] Another contributing factor is the wavevector dependent Berry curvature that leads to changes of lattice orbital angular momentum away from the K points. In light of these considerations, observing both σ+ and σ- from the same valley is not completely surprising.

The switching of photon helicity has been frequently used as an indicator for monitoring intervalley scattering of photo-excited charge carriers.[21, 22] In 1L-MX$_2$ switching of the hole valley pseudospin necessitates a concomitant spin flip, which can only occur in the presence of both time reversal symmetry breaking and large quasi-momentum transfer, e.g. an atomically sharp magnetic defect.[21] However in the Raman process of our experiment the photon helicity is switched involving neither of the two processes. Our data reflect the fact that the degree of photon circular polarization and the degree of electron valley polarization are not equivalent. The quantitative relation between the two parameters requires further experimental and theoretical investigation.

The observations in Fig.2 (b) & (c) suggest that the helicity selection rules are robust, and do not rely on the spin splitting of valence bands (present in inversion-symmetry-broken 1L but absent in inversion-symmetric 2L) nor the orbital character of specific electron states (as the helicity selection rules show no dependence on the incident photon energy which determines the electron-hole pair energy). Instead, this effect is generic and can be explained by the symmetry of relevant phonon modes. Monolayer $MoS_2$ is invariant under the 12 symmetry operations in the $D_{3h}$ point group.[30, 35] As summarized in Table 1, the zone center optical phonons in Fig. 1(a) transform according to the irreducible representations of the group: $E'$ for IMC and $A_1'$ for OC are even under the mirror reflection $\sigma_h$ about the Mo atomic plane, while $E''$ for IC and $A_2''$ for OMC are odd. Bilayer $MoS_2$ is symmetric under inversion and the symmetry operations form the $D_{3d}$ point group. The optical phonons in Fig. 1(a) of the two layers couple to form, under inversion operation of $D_{3d}$, symmetric modes ($E_g$ for IC and IMC, $A_{1g}$ for OC and OMC) and anti-symmetric modes ($E_u$ for IC and IMC, $A_{2u}$ for OC and OMC). Coupling of the acoustic phonons forms the even $E_g$ shear mode and $A_{1g}$ breathing mode (Fig. 1(b)), while the bilayer acoustic LA/TA and ZA are odd under inversion and transform as $E_u$ and $A_{2u}$ representations. Going further, it turns out the symmetry point group is $D_{3h}$ for all odd layers, and $D_{3d}$ for all even layers (note that in this discussion we only consider 2Hc stacking[30]). Details of the symmetry representations of all zone center phonon modes are summarized in Table 1. For the centrosymmetric 2Hc stacking bulk

MoS$_2$, the unit cell consists of 2 MX$_2$ units with 6 atoms, and the symmetry space group is the non-symmorphic D$^4_{6h}$ (P6$_3$/*mmc*). The acoustic phonons are odd under inversion while the shear and breathing modes are even, and the IC, OC, IMC, and OMC each have one even and one odd.

The observed IMC and OC modes in Fig. 2(b) thus transform according to the $E_{2g}$/ $E'$/$E_g$ and $A_{1g}$/ $A_1'$/$A_{1g}$ representations for bulk/odd/even number of layers. The photon energies used in the experiment are much larger than the phonon energies. One can use tabulated Raman tensors to find polarization state of the scattered photon.[47] The Raman cross section is given by $A \sum_{j=1}^{d}|\langle\varepsilon_o|R_j|\varepsilon_i\rangle|^2$ where $A$ is a constant, $R_j$'s are Raman tensors, $\varepsilon_i$ and $\varepsilon_o$ are polarization states of the incoming and outgoing light. For OC modes, the Raman tensor for $A_{1g}$/ $A_1'$/$A_{1g}$ representations is the same, which is given by $\begin{bmatrix} a & 0 & 0 \\ 0 & a & 0 \\ 0 & 0 & b \end{bmatrix}$. For same circular-polarization incident and outgoing light, e.g., $\sigma_i = \sigma_o = \frac{1}{\sqrt{2}}\begin{bmatrix} 1 \\ -i \\ 0 \end{bmatrix}$, $\langle\varepsilon_o|R_j|\varepsilon_i\rangle = \sigma_o^\dagger R_j \sigma_i = a$. On the contrary, if $\sigma_i$ and $\sigma_o$ have opposite helicity, $\sigma_o^\dagger R_j \sigma_i = 0$. This observation is consistent with Fig. 2(c) where the OC modes have the same helicity as that of the incident and the Rayleigh scattered light. For the IMC modes, the Raman tensors are $\begin{bmatrix} 0 & d & 0 \\ d & 0 & 0 \\ 0 & 0 & 0 \end{bmatrix}$ and $\begin{bmatrix} d & 0 & 0 \\ 0 & -d & 0 \\ 0 & 0 & 0 \end{bmatrix}$ for both $E_{2g}$ and $E'$ representations; and are $\begin{bmatrix} 0 & -c & -d \\ -c & 0 & 0 \\ -d & 0 & 0 \end{bmatrix}$ and $\begin{bmatrix} c & 0 & 0 \\ 0 & -c & d \\ 0 & d & 0 \end{bmatrix}$ for the $E_g$ representation. Explicit calculations show that for these Raman tensors $\sigma_o^\dagger R_j \sigma_i$ is zero (nonzero) for same (opposite) $\sigma_i$ and $\sigma_o$ helicities. This similarly explains why in Fig. 2(b) the IMC modes are absent in the σ+ polarized spectra when the incident light is σ+ polarized.

The above discussion of $MoS_2$ OC and IMC modes is based on symmetry and is thus applicable to the family of semiconducting TMDC atomic layers with the same crystal structure. We further studied the OC and IMC modes in $WS_2$, $WSe_2$ and $MoSe_2$ atomic layers as shown in Figure 3. As expected all the OC Raman scattering signals have the same helicity as the incident while the IMC modes occur only in spectra of opposite helicity. The OC modes of $MoS_2$ and $WS_2$ have similar energies 410±10 cm$^{-1}$,[31-36, 48, 49] reflecting the fact that they both involve only sulfur atoms and that there is only a slight difference between the bond strength (spring constant) in the two materials. Similarly OC modes in $MoSe_2$ and $WSe_2$ have energies of 245±5 cm$^{-1}$ which are lower due to the larger mass of selenium atoms.[35-40] For the IMC modes, the inverse of the reduced mass is given by the sum of the inverse mass of one metal atom and that of two chalcogen atoms. Thus the light mass of sulfur atoms (2×32) make IMC in $MoS_2$ and $WS_2$ have higher energy (370±15 cm$^{-1}$) than in $MoSe_2$ (Mo: 96; 288 cm$^{-1}$) and $WSe_2$ (W: 184; Se: 2×79; 250cm$^{-1}$). In particular, in monolayer $WSe_2$, the IMC mode and the OC become accidentally degenerate.

The observed helicity selection rules are useful for assigning and resolving Raman bands. In literature typically polarized Raman scattering of TMDC atomic layers is performed with linearly polarized light. While the OC mode only shows up in parallel polarization, IMC is allowed for both parallel and cross polarization. This can sometimes become problematic for resolving Raman modes. For instance, in the Raman spectra of $WS_2$ (Fig. 3(a)), the IMC modes (~360 cm$^{-1}$) and 2LA modes (~358 cm$^{-1}$) are almost degenerate, making them difficult to resolve with linearly polarized light. With circularly polarized light, one can clearly distinguish contributions from 2LA and IMC, since the latter is forbidden in σ+. Another typical example is the OC and IMC modes of WSe$_2$ that have almost the same energy in the monolayer limit. It becomes a bit difficult to trace the evolution of these two modes by linearly polarized measurements. In fact, in literature there have been conflicting conclusions about assignment of these Raman peaks.[35-38, 50] Our helicity-resolved Raman scattering in Fig. 3(b) clearly resolves the two Raman modes despite that their energies are very close, since one only shows up in σ+ spectra while the other only in σ- spectra. Besides, by checking the Raman selection rule, we can clearly distinguish IMC and OC from the 2LA band, which is located in the range of 260-270 cm$^{-1}$ and has caused some confusion.

In the helicity-resolved Raman scattering measurements of MoSe$_2$ samples, the OC modes split when the thickness increases to trilayer. The peak splitting is about 3 cm$^{-1}$ for 3L-MoSe$_2$ and 2 cm$^{-1}$ for 4L-MoSe$_2$, respectively, consistent with previous studies.[36] This is understood by the above group theory analysis: for $N = 3$, there are $(N + 1)/2 = 2$ OC normal modes with $A_1'$ symmetry; for $N = 4$ there are $N/2 = 2$ OC normal modes with $A_{1g}$ symmetry. This splitting is known as Davydov splitting due to interlayer interactions.[49] Interestingly, while theoretically such splitting should also occur in other multilayer TMDCs as well as for the IMC modes, it is only experimentally observed in MoSe$_2$ OC lattice vibration. We speculate that this is because the splitting is not large enough compared with the line-width of the relevant phonon bands.

The helicity selection rules are anticipated to apply as well to the shear and breathing modes in few layer TMDCs. From Table 1, the Faraday-geometry Raman-active breathing modes are of $A_{1g}/A_1'$ symmetry, and the shear modes have $E_g/E'$ symmetry in even/odd layers of TMDC. This is the same as the OC and IMC modes respectively. Figure 4 shows our experimental data in the four TMDC atomic layers. Indeed, the breathing mode preserves photon helicity while the shear mode reverses photon helicity.

Our Raman spectra in Fig. 4 are in agreement with recent studies of $MoS_2$ and $WSe_2$ using linearly polarized light,[32, 35] and provide the first measurement of these low-lying modes in $MoSe_2$ and $WS_2$. The spectra are strongly layer thickness dependent, with the shear (breathing) mode stiffening (softening) with increasing number of layers, and absent in monolayer TMDC as expected. This sensitive dependence, similar to that observed in multi-layer graphene,[51-53] has been interpreted by a linear chain model and provides a sensitive fingerprint for TMDC atomic layer number identification.[32, 35] The advantage of helicity-resolved measurement can be seen in 3L TMDCs where the B and S modes have very similar energies. The capability to separately resolve S and B modes using helicity-dependent Raman provides higher accuracy in distinguishing the subtle mode energy differences, as compared with unpolarized or linearly polarized measurements[32, 35] in which the B (partially) overlaps with the S mode and can only be analyzed via multi-peak fitting. We list all our measured mode energies in Table 2.

In summary, we studied helicity-resolved Raman scattering of the TMDC atomic layers. The switching of photon angular momentum by zone-center optical phonons is interpreted as a result of phonon symmetry, instead of intervalley scattering and spin flip, providing new insights into the relation between photon helicity and valley pumping. The helicity selection rule is found to be robust and generic for different modes in four different materials, offering a very useful tool for TMDC atomic layer characterization. The experimental method, as compared with unpolarized or linearly polarized Raman scattering, is more advantageous in distinguishing and assigning phonon modes, as well as providing more accurate measurements especially in the presence of accidental degeneracy. We further anticipate that, from the generic symmetry considerations presented here, the helicity-resolved Raman spectroscopy is applicable to excitations in other materials, and will be a powerful tool for probing breathing modes, shear modes, layer stacking etc. in all 2D layered systems.

**Methods**

**Sample Preparation**. MoS$_2$ (SPI Supplies, Inc., USA), MoSe$_2$ (2d semiconductors, Inc., USA) and WSe$_2$ (Nanoscience Instruments, Inc. USA) samples were fabricated by mechanically exfoliating the commercial flakes directly on 280 nm SiO$_2$/Si substrates. WS$_2$ samples were grown by chemical vapor deposition on polished sapphire substrates. The precursors are sulfur (Sigma-Aldrich, 1 g) and WO$_3$ powders (Sigma-Aldrich, 2.0 g). Sapphire substrates were cleaned by 10 min acetone and isopropanol sonication, followed by oxygen annealing (500 sccm) at 1050 ℃ for 1h. The growth was performed in a 1 inch diameter quartz tube flowing ultra-high purity argon (200 sccm) and hydrogen (10 sccm) at atmospheric pressure. During the growth, WO$_3$ and sapphire substrates were loaded into separate quartz crucibles and positioned at the center and downstream end of the furnace, respectively. The sulfur powder was located upstream of the WO$_3$ and was independently heated. The furnace and the sulfur are heated to 800 ℃ and 200 ℃ respectively in 40 min, followed by maximum cooling achievable with furnace opening. The thickness of flakes is determined by optical contrast as well as photoluminescence and Raman scattering.

**Optical Measurement**. The helicity-resolved Raman measurement was performed with a micro-Raman Spectrometer (Horiba T64000) equipped with a liquid nitrogen cooled CCD. The excitation laser was first guided through a vertical linear polarizer followed by a broadband quarter-wave plate (Fresnel Rhomb Retarder) to achieve σ+ circular polarization. The circular polarization of the excitation light was confirmed at the sample position. The back-scattered Raman signal going through the same quarter-wave plate was collected and analyzed with a broadband half-wave plate and a linear polarizer. Rotation of the half-wave plate at different angles enables us to obtain detailed information on the helicity of the scattered light. We used three different laser sources for the wavelength dependent measurements: Argon laser (488 nm), Nd:YAG solid state laser (532 nm) and dye laser (605nm-645nm, dye: Kiton Red/LC6200). To achieve a high quality laser line for the low wavenumber phonon mode measurements down to 10 cm$^{-1}$, the laser was first reflected by a holographic diffraction grating and followed by a tunable plasma line filter.

**Contributions**

S.-Y.C. and J.Y. conceived the optical experiment. S.-Y.C. fabricated the $MoS_2$, $MoSe_2$ and $WSe_2$ samples, and performed the helicity-resolved luminescence and Raman scattering measurements. M.S.F and C.Z. conceived the CVD growth of $WS_2$ on sapphire, and C.Z. carried out the growth. S.-Y.C. and J.Y. co-wrote the paper. All authors discussed the results, edited, commented and agreed on the manuscript.

**Acknowledgements**

This work is supported by the University of Massachusetts Amherst and the National Science Foundation Center for Hierarchical Manufacturing (CMMI-1025020). M.S.F and C.Z. acknowledge support from ARC (DP150103837 and FL120100038). C.Z. acknowledges support from ARC DECRA (DE140101555).

# References

1. Novoselov, K. S.; Geim, A. K.; Morozov, S. V.; Jiang, D.; Zhang, Y.; Dubonos, S. V.; Grigorieva, I. V.; Firsov, A. A. *Science* **2004,** 306, 666-669.

2. Geim, A. K.; Grigorieva, I. V. *Nature* **2013,** 499, 419-425.

3. Butler, S. Z.; Hollen, S. M.; Cao, L.; Cui, Y.; Gupta, J. A.; Gutiérrez, H. R.; Heinz, T. F.; Hong, S. S.; Huang, J.; Ismach, A. F.; Johnston-Halperin, E.; Kuno, M.; Plashnitsa, V. V.; Robinson, R. D.; Ruoff, R. S.; Salahuddin, S.; Shan, J.; Shi, L.; Spencer, M. G.; Terrones, M.; Windl, W.; Goldberger, J. E. *ACS Nano* **2013,** 7, 2898-2926.

4. Wang, Q. H.; Kalantar-Zadeh, K.; Kis, A.; Coleman, J. N.; Strano, M. S. *Nature nanotechnology* **2012,** 7, 699-712.

5. Novoselov, K. S.; Jiang, D.; Schedin, F.; Booth, T. J.; Khotkevich, V. V.; Morozov, S. V.; Geim, A. K. *Proceedings of the National Academy of Sciences of the United States of America* **2005,** 102, 10451-10453.

6. Splendiani, A.; Sun, L.; Zhang, Y.; Li, T.; Kim, J.; Chim, C. Y.; Galli, G.; Wang, F. *Nano letters* **2010,** 10, 1271-1275.

7. Mak, K. F.; Lee, C.; Hone, J.; Shan, J.; Heinz, T. F. *Physical Review Letters* **2010,** 105, 136805.

8. Ye, Z.; Cao, T.; O'Brien, K.; Zhu, H.; Yin, X.; Wang, Y.; Louie, S. G.; Zhang, X. *Nature* **2014,** 513, 214-218.

9. Chernikov, A.; Berkelbach, T. C.; Hill, H. M.; Rigosi, A.; Li, Y.; Aslan, O. B.; Reichman, D. R.; Hybertsen, M. S.; Heinz, T. F. *Physical Review Letters* **2014,** 113, 076802.

10. He, K.; Kumar, N.; Zhao, L.; Wang, Z.; Mak, K. F.; Zhao, H.; Shan, J. *Physical Review Letters* **2014,** 113, 026803.

11. Mak, K. F.; He, K.; Lee, C.; Lee, G. H.; Hone, J.; Heinz, T. F.; Shan, J. *Nature materials* **2013,** 12, 207-211.

12. Cheng, R.; Li, D.; Zhou, H.; Wang, C.; Yin, A.; Jiang, S.; Liu, Y.; Chen, Y.; Huang, Y.; Duan, X. *Nano letters* **2014,** 14, 5590-5597.

13. Ross, J. S.; Klement, P.; Jones, A. M.; Ghimire, N. J.; Yan, J.; Mandrus, D. G.; Taniguchi, T.; Watanabe, K.; Kitamura, K.; Yao, W.; Cobden, D. H.; Xu, X. *Nature nanotechnology* **2014,** 9, 268-272.

14. Xia, F.; Wang, H.; Xiao, D.; Dubey, M.; Ramasubramaniam, A. *Nature Photonics* **2014,** 8, 899-907.

15. Radisavljevic, B.; Radenovic, A.; Brivio, J.; Giacometti, V.; Kis, A. *Nature nanotechnology* **2011,** 6, 147-150.

16. Radisavljevic, B.; Kis, A. *Nature materials* **2013,** 12, 815-820.

17. Chuang, H. J.; Tan, X.; Ghimire, N. J.; Perera, M. M.; Chamlagain, B.; Cheng, M. M.; Yan, J.; Mandrus, D.; Tomanek, D.; Zhou, Z. *Nano letters* **2014,** 14, 3594-3601.

18. Baugher, B. W.; Churchill, H. O.; Yang, Y.; Jarillo-Herrero, P. *Nano letters* **2013,** 13, 4212-4216.

19. Bao, W.; Cai, X.; Kim, D.; Sridhara, K.; Fuhrer, M. S. *Applied Physics Letters* **2013,** 102, 042104.

20. Xiao, D.; Liu, G.-B.; Feng, W.; Xu, X.; Yao, W. *Physical Review Letters* **2012,** 108, 196802.

21. Mak, K. F.; He, K.; Shan, J.; Heinz, T. F. *Nature nanotechnology* **2012,** 7, 494-498.

22. Zeng, H.; Dai, J.; Yao, W.; Xiao, D.; Cui, X. *Nature nanotechnology* **2012,** 7, 490-493.

23. Cao, T.; Wang, G.; Han, W.; Ye, H.; Zhu, C.; Shi, J.; Niu, Q.; Tan, P.; Wang, E.; Liu, B.; Feng, J. *Nature communications* **2012,** 3, 887.

24. Jones, A. M.; Yu, H.; Ghimire, N. J.; Wu, S.; Aivazian, G.; Ross, J. S.; Zhao, B.; Yan, J.; Mandrus, D. G.; Xiao, D.; Yao, W.; Xu, X. *Nature nanotechnology* **2013,** 8, 634-638.

25. Mak, K. F.; McGill, K. L.; Park, J.; McEuen, P. L. *Science* **2014,** 344, 1489-1492.

26. Xiao, D.; Yao, W.; Niu, Q. *Physical Review Letters* **2007,** 99, 236809.

27. Rycerz, A.; Tworzydło, J.; Beenakker, C. W. J. *Nature Physics* **2007,** 3, 172-175.

28. Malard, L. M.; Pimenta, M. A.; Dresselhaus, G.; Dresselhaus, M. S. *Physics Reports* **2009,** 473, 51-87.

29. Reich, S.; Thomsen, C. *Philosophical transactions. Series A, Mathematical, physical, and engineering sciences* **2004,** 362, 2271-2288.

30. Ribeiro-Soares, J.; Almeida, R. M.; Barros, E. B.; Araujo, P. T.; Dresselhaus, M. S.; Cançado, L. G.; Jorio, A. *Physical Review B* **2014,** 90, 115438.

31. Lee, C.; Yan, H.; Brus, L. E.; Heinz, T. F.; Hone, J.; Ryu, S. *ACS Nano* **2010,** 4, 2695-2700.

32. Zhang, X.; Han, W.; Wu, J.; Milana, S.; Lu, Y.; Li, Q.; Ferrari, A.; Tan, P. *Physical Review B* **2013,** 87, 115413.

33. Berkdemir, A.; Gutiérrez, H. R.; Botello-Méndez, A. R.; Perea-López, N.; Elías, A. L.; Chia, C.-I.; Wang, B.; Crespi, V. H.; López-Urías, F.; Charlier, J.-C.; Terrones, H.; Terrones, M. *Scientific reports* **2013,** 3, 1755.

34. Mitioglu, A. A.; Plochocka, P.; Deligeorgis, G.; Anghel, S.; Kulyuk, L.; Maude, D. K. *Physical Review B* **2014,** 89, 245442.

35. Zhao, Y.; Luo, X.; Li, H.; Zhang, J.; Araujo, P. T.; Gan, C. K.; Wu, J.; Zhang, H.; Quek, S. Y.; Dresselhaus, M. S.; Xiong, Q. *Nano letters* **2013,** 13, 1007-1015.

36. Tonndorf, P.; Schmidt, R.; Böttger, P.; Zhang, X.; Börner, J.; Liebig, A.; Albrecht, M.; Kloc, C.; Gordan, O.; Zahn, D. R. T.; Vasconcellos, S. M. d.; Bratschitsch, R. *Optics Express* **2013,** 21, 4908-4916.

37. Terrones, H.; Del Corro, E.; Feng, S.; Poumirol, J. M.; Rhodes, D.; Smirnov, D.; Pradhan, N. R.; Lin, Z.; Nguyen, M. A.; Elias, A. L.; Mallouk, T. E.; Balicas, L.; Pimenta, M. A.; Terrones, M. *Scientific reports* **2014,** 4, 4215.

38. Zhao, W.; Ghorannevis, Z.; Amara, K. K.; Pang, J. R.; Toh, M.; Zhang, X.; Kloc, C.; Tan, P. H.; Eda, G. *Nanoscale* **2013,** 5, 9677-9683.

39. Luo, X.; Zhao, Y.; Zhang, J.; Toh, M.; Kloc, C.; Xiong, Q.; Quek, S. Y. *Physical Review B* **2013,** 88, 195313.

40. Late, D. J.; Shirodkar, S. N.; Waghmare, U. V.; Dravid, V. P.; Rao, C. N. *Chemphyschem : a European journal of chemical physics and physical chemistry* **2014,** 15, 1592-1598.

41. Kossacki, P.; Faugeras, C.; Kühne, M.; Orlita, M.; Mahmood, A.; Dujardin, E.; Nair, R. R.; Geim, A. K.; Potemski, M. *Physical Review B* **2012,** 86, 205431.

42. Mialitsin, A.; Kim, J.; Kremer, R.; Blumberg, G. *Physical Review B* **2009,** 79, 064503.

43. Qazilbash, M.; Koitzsch, A.; Dennis, B.; Gozar, A.; Balci, H.; Kendziora, C.; Greene, R.; Blumberg, G. *Physical Review B* **2005,** 72, 214510.

44. Cappelluti, E.; Roldán, R.; Silva-Guillén, J. A.; Ordejón, P.; Guinea, F. *Physical Review B* **2013,** 88, 075409.

45. Liu, G.-B.; Shan, W.-Y.; Yao, Y.; Yao, W.; Xiao, D. *Physical Review B* **2013,** 88, 085433.

46. Li, T.; Galli, G. *J. Phys. Chem. C* **2007,** 111, 16192-16196.

47. Loudon, R. *Advances in Physics* **1964,** 13, 423-482.

48. Li, H.; Zhang, Q.; Yap, C. C. R.; Tay, B. K.; Edwin, T. H. T.; Olivier, A.; Baillargeat, D. *Advanced Functional Materials* **2012,** 22, 1385-1390.

49. Molina-Sánchez, A.; Wirtz, L. *Physical Review B* **2011,** 84, 155413.

50. Huang, J.-K.; Pu, J.; Hsu, C.-L.; Chiu, M.-H.; Juang, Z.-Y.; Chang, Y.-H.; Chang, W.-H.; Iwasa, Y.; Takenobu, T.; Li, L.-J. *ACS Nano* **2014,** 8, 923-930.

51. Tan, P. H.; Han, W. P.; Zhao, W. J.; Wu, Z. H.; Chang, K.; Wang, H.; Wang, Y. F.; Bonini, N.; Marzari, N.; Pugno, N.; Savini, G.; Lombardo, A.; Ferrari, A. C. *Nature materials* **2012,** 11, 294-300.

52. Lui, C. H.; Heinz, T. F. *Physical Review B* **2013,** 87, 121404.

53. Lui, C. H.; Ye, Z.; Keiser, C.; Xiao, X.; He, R. *Nano letters* **2014,** 14, 4615-21.

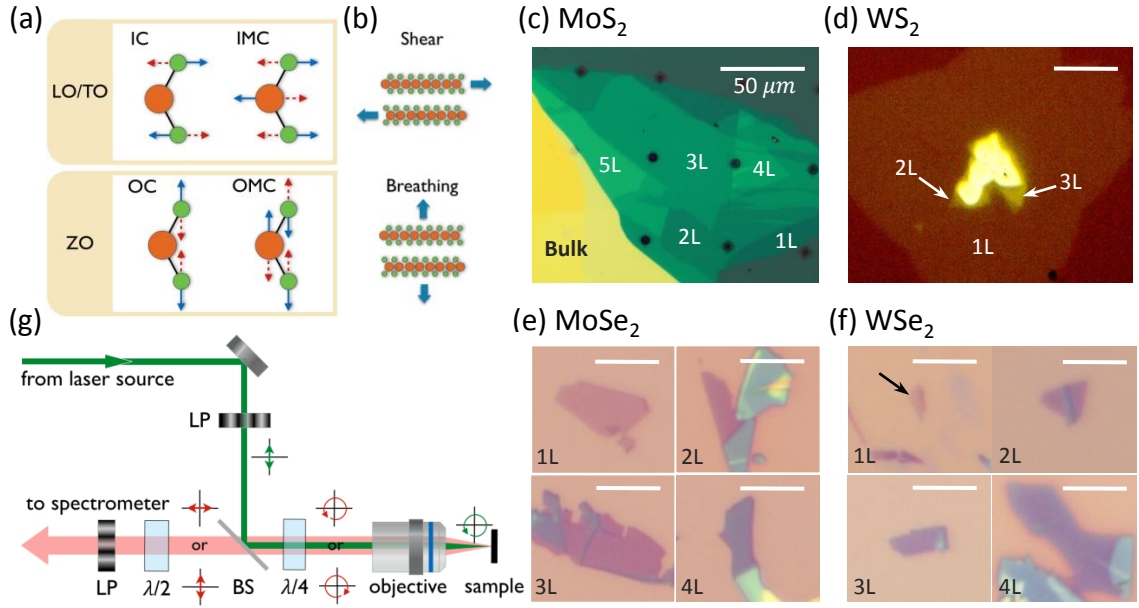

**Figure 1**. (**a**) Optical phonons at the Brillouin zone center of monolayer TMDCs. (**b**) Interlayer shear and breathing vibrational modes in bilayer TMDCs. (**c-f**) Optical microscope images for atomic layers of (**c**) exfoliated $MoS_2$, (**d**) CVD $WS_2$, (**e**) exfoliated $MoSe_2$ and (**f**) exfoliated $WSe_2$ samples. The scale bars are 5 $\mu m$ except for panel (**c**). (**g**) Experimental setup for helicity-resolved micro-Raman spectroscopy. The green (red) path is for incident (scattered) light.

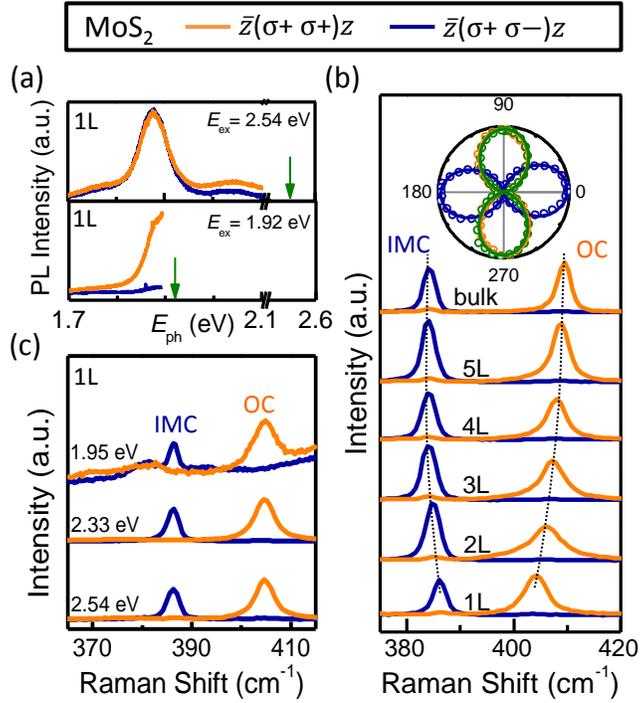

**Figure 2.** (**a**) The helicity-resolved photoluminescence of 1L-MoS$_2$ at 80 K with excitation photon energy at 2.54 eV (upper) and 1.92 eV (lower). (**b**) Helicity-resolved Raman spectra of 1L-5L and bulk MoS$_2$. The excitation wavelength is 488 nm with $\sigma+$ polarization. Inset: normalized angular dependence of the Rayleigh (green), IMC (blue) and OC phonon (orange) scattering intensities for 1L-MoS$_2$. (**c**) Raman scattering of 1L-MoS$_2$ with various excitation photon energies.

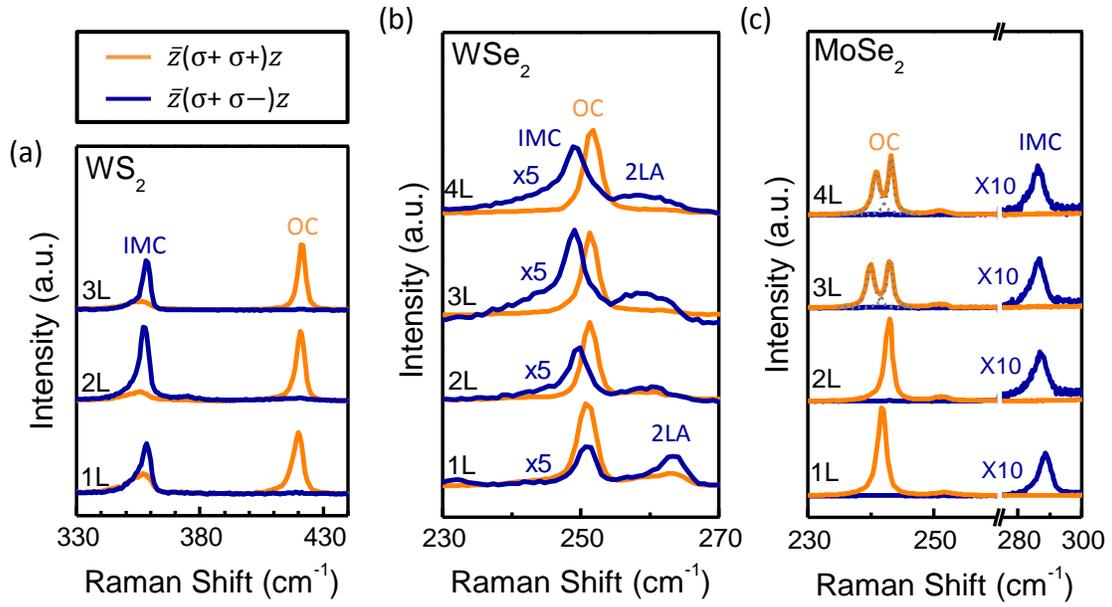

**Figure 3.** Helicity-resolved Raman scattering in (**a**) WS$_2$ (**b**) WSe$_2$ and (**c**) MoSe$_2$ with various thickness. All spectra are normalized by the peak intensity of OC modes and vertically shifted for clarity.

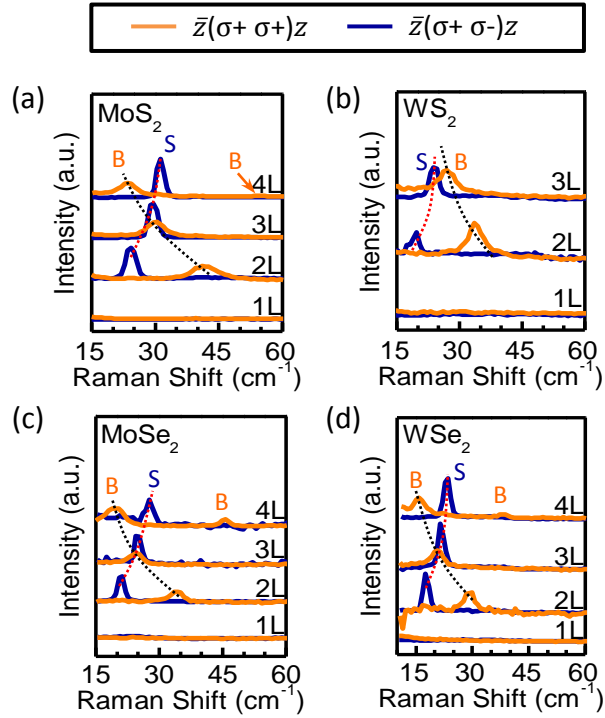

**Figure 4.** The helicity-resolved shear (S) and breathing (B) modes for (**a**) $MoS_2$ (**b**) $WS_2$ (**c**) $MoSe_2$ and (**d**) $WSe_2$. The red and black dashed curves are guides to the eye for the evolution of the shear and breathing modes respectively with atomic layer thickness.

**Table 1.** Symmetry representations for phonon modes in bulk and few layer TMDCs.

| # of Layer | Sym. Grp. | $\sigma_h$/i Sym.* | LA/TA | ZA | LO/TO | | | ZO | | |
|---|---|---|---|---|---|---|---|---|---|---|
| | | | | | IC | IMC | Shear | OC | OMC | Breathing |
| 1 | $D_{3h}$ | + | $E'$ | - | - | 1 $E'$ | - | 1 $A'_1$ | - | - |
| | | − | - | $A''_2$ | 1 $E''$ | - | - | - | 1 $A''_2$ | - |
| 2 | $D_{3d}$ | + | - | - | 1 $E_g$ | 1 $E_g$ | 1 $E_g$ | 1 $A_{1g}$ | 1 $A_{1g}$ | 1 $A_{1g}$ |
| | | − | $E_u$ | $A_{2u}$ | 1 $E_u$ | 1 $E_u$ | - | 1 $A_{2u}$ | 1 $A_{2u}$ | - |
| odd N | $D_{3h}$ | + | $E'$ | - | (N-1)/2 $E'$ | (N+1)/2 $E'$ | (N-1)/2 $E'$ | (N+1)/2 $A'_1$ | (N-1)/2 $A'_1$ | (N-1)/2 $A'_1$ |
| | | − | - | $A''_2$ | (N+1)/2 $E''$ | (N-1)/2 $E''$ | (N-1)/2 $E''$ | (N-1)/2 $A''_2$ | (N+1)/2 $A''_2$ | (N-1)/2 $A''_2$ |
| even N | $D_{3d}$ | + | - | - | N/2 $E_g$ | N/2 $E_g$ | N/2 $E_g$ | N/2 $A_{1g}$ | N/2 $A_{1g}$ | N/2 $A_{1g}$ |
| | | − | $E_u$ | $A_{2u}$ | N/2 $E_u$ | N/2 $E_u$ | (N-2)/2 $E_u$ | N/2 $A_{2u}$ | N/2 $A_{2u}$ | (N-2)/2 $A_{2u}$ |
| bulk | $D^4_{6h}$ | + | - | - | 1 $E_{1g}$ | 1 $E_{2g}$ | 1 $E_{2g}$ | 1 $A_{1g}$ | 1 $B_{2g}$ | 1 $B_{2g}$ |
| | | − | $E_{1u}$ | $A_{2u}$ | 1 $E_{2u}$ | 1 $E_{1u}$ | - | 1 $B_{1u}$ | 1 $A_{2u}$ | - |

*The third column specifies whether the mode is even or odd under horizontal mirror plane reflection (inversion) for odd number of layers (even number of layers and bulk).

**Table 2.** Extracted mode energies in cm$^{-1}$ (1 meV = 8.07 cm$^{-1}$) for the shear, breathing, IMC and OC modes of TMDCs.

|  |  | 1L | 2L | 3L | 4L |
|---|---|---|---|---|---|
| MoS$_2$ | S |  | 24.2 | 29.6 | 31.2 |
|  | B |  | 41.6 | 30.1 | 23.5/54.1 |
|  | IMC | 386.3 | 384.9 | 384.2 | 384.1 |
|  | OC | 404.1 | 406.1 | 407.2 | 408.1 |
| WS$_2$ | S |  | 19.6 | 24.2 | - |
|  | B |  | 33.8 | 27.0 | - |
|  | IMC | 359.0 | 358.3 | 358.0 | - |
|  | OC | 420.4 | 420.8 | 421.2 | - |
| MoSe$_2$ | S |  | 21.0 | 24.9 | 26.4 |
|  | B |  | 34.3 | 24.5 | 18.2/45.4 |
|  | IMC | 288.6 | 287.1 | 286.5 | 286.4 |
|  | OC | 241.8 | 242.8 | 239.9/243.0 | 240.9/243.2 |
| WSe$_2$ | S |  | 17.7 | 21.6 | 23.2 |
|  | B |  | 29.1 | 20.5 | 15.7/38.0 |
|  | IMC | 250.8 | 249.6 | 249.1 | 248.9 |
|  | OC | 250.8 | 251.2 | 251.4 | 251.6 |